\documentclass[a4paper,10pt,onecolumn]{article}

\usepackage[]{achemso}
\usepackage[para]{threeparttable}
\usepackage{amsmath}
\usepackage{amsfonts}
\usepackage{amssymb}
\usepackage{bm}
\usepackage{scrtime}[24hr]
\usepackage[usenames]{color}
\usepackage{rcs} 
\usepackage[dvips]{graphicx}

\newcommand{\degc}{\ensuremath{^{\circ}}\mathrm{C}}
\newcommand{\stequil}[1]{\stackrel{#1}{\rightleftharpoons}}

\def\mys#1{{\mbox{\scriptsize{#1}}}}    

\def\R0{R}                      
\def\RM{r}                      
\def\RH{r_{\mys{h}}}            
\def\VM{v_{\mys{m}}}            

\def\Z0{Z_{\mys{0}}}            

\def\z{z}                       
\def\CH{\chi}                   

\def\NP{N}                      
\def\MP{M}                      
\def\ZP{Z_{\mys{P}}}            
\def\ZN{Z_{\mys{N}}}            
\def\ZU{Z_{\mys{U}}}            

\def\GP{\Delta g_{\mys{P}}}            
\def\GN{\Delta g_{\mys{N}}}            
\def\GU{\Delta g_{\mys{U}}}            

\def\DMU{\Delta \mu}            

\def\GM{u_{\mys{el}}}           
\def\GMZ{U_{\mys{el}}}           

\def\KP{K_{\mys{P}}}            
\def\KN{K_{\mys{N}}}            
\def\KU{K_{\mys{U}}}            

\def\vol{\varphi_{\mys{m}}}     
\def\volc{\varphi_{\mys{c}}}    
\def\kb{k_{\mys{B}}}            
\def\b{\beta}                   
\def\lb{\lambda_{\mys{B}}}      
\def\ex{\mathrm{exp}}           
\def\pot{\phi}                  
\def\PH{\Phi}                   
\def\Ps{\Phi^{\star}}           
\def\Zs{Z_{\mys{0}}^{\star}}    
\def\alp{\alpha}                
\def\vm{v_{\mys{m}}}            

\renewcommand{\textfraction}{0.0}
\renewcommand{\topfraction}{1.0}
\renewcommand{\bottomfraction}{1.0}
\begin{document}
\unitlength 1cm
\renewcommand{\textfraction}{0.0}
\renewcommand{\topfraction}{1.0}
\renewcommand{\bottomfraction}{1.0}
\Roman{section}

\title{\bf{Electrostatic charging of non-polar colloids by reverse micelles}}

\author{G. Seth Roberts, Rodrigo Sanchez, Roger Kemp,
Tiffany Wood, \\ and Paul Bartlett\footnote{Corresponding author. Email: p.bartlett@bristol.ac.uk.}\
\\{\em School of Chemistry, University of Bristol, Bristol BS8
1TS, UK.}
}

\RCS $Revision: 1.0 $ 
\RCSdate $Date: 2007/12/14 11:53:12 $ 

\date{\small Version [\RCSRevision]: \RCSDate, \RCSTime.} 

\maketitle
\begin{abstract}

Colloids dispersed in a non-polar solvent become charged when reverse micelles are added. We study the charge of individual sterically-stabilized poly(methyl methacrylate) spheres dispersed in micellar solutions of the surfactants sodium bis(2-ethyl 1-hexyl) sulfosuccinate [AOT], zirconyl 2-ethyl hexanoate [Zr(Oct)$_{2}$], and a copolymer of poly(12-hydroxystearic acid)--poly(methyl methacrylate) [PHSA-PMMA]. Although the sign of the particle charge is positive for Zr(Oct)$_{2}$, negative for AOT, and essentially neutral for PHSA-PMMA the different micellar systems display a number of common features. In particular, we demonstrate that, over a wide range of concentrations, the colloid charge is independent of the number of micelles added and scales linearly with the colloid size. A simple thermodynamic model, in which the particle charge is generated by the competitive adsorption of both positive and negative micelles, is in good agreement with the experimental data.

\end{abstract}

\section{Introduction} \label{sec-intro}


Interactions between surface active materials and non-polar suspensions of colloidal particles play a key role in many technologically important processes. It has been recognized for at least fifty years\cite{4951} that adding surfactants to a non-polar suspension frequently results in particle charging. This phenomenon is important in many practical situations including the formulation of electrophoretic image displays \cite{3762,3772,3444}, electrorheological fluids \cite{5052},  air-borne drug delivery systems \cite{4576}, drop-on-demand ink jet printing\cite{Morrison-868}, liquid electrostatic developers \cite{3480,4939}, and liquid detergents \cite{4436} together with the prevention of asphaltene deposits in crude oil processing \cite{4949}, colloidal stabilization in supercritical CO$_{2}$ \cite{3775}, flow electrification in petroleum handling \cite{5051,4948}, and the synthesis of new materials \cite{3738,3958}. It is surprising therefore that, given its technological significance, the mechanism of charging in solvents of low permittivity is not well understood\cite{Morrison-868}.

A non-polar solvent is distinguished from a polar solvent by a low relative dielectric constant, $\epsilon_{r}$, of typically around 2 -- 5.  The thermodynamics of charging in a liquid is controlled by the Bjerrum length $\lb$,
\begin{equation}\label{eqn-bjerrum-length}
    \lb =  \frac{e^{2}}{4 \pi \epsilon_{0} \epsilon_{r} \kb T},
\end{equation}
which is the characteristic separation between two ions at which their Coulombic interactions are exactly balanced by the thermal energy ($\kb T$). Here $e$ is the elementary charge and $\epsilon_{0}$ is the vacuum permittivity. In water, where $\epsilon_{r} = 80$ at 22 $\degc$,  $\lb$ is only 0.7 nm whilst for a typical oil such as dodecane ($\epsilon_{r} = 2$) the Bjerrum length is some 40 times larger at $\lb = 28.3$ nm. The large Bjerrum length in a non-polar environment
has two important consequences for colloids. First, the concentration of molecular ions
is extremely small because the solvation energy of an ion scales\cite{4981} as $\lb / 2a$  where $a$ is the ionic radius. Because of the practical absence of charge carriers in an oil, screening of electrostatic interactions is negligible and charge interactions are extremely long-ranged. To demonstrate this, consider the dissociation of the symmetric monovalent electrolyte A$^{+}$A$^{-}$,
\begin{equation}\label{eqn-electrolyte-dissociation}
    \textrm{A}^{+} \textrm{A}^{-} \rightleftharpoons \textrm{A}^{+} + \textrm{A}^{-}.
\end{equation}
Applying the law of mass action to this chemical equilibrium yields an expression for the total number of free ions per unit volume,
\begin{equation}\label{eqn-rho-ions}
    \rho_{\textrm{ion}} = \sqrt{\frac{3\rho}{\pi a^{3}} \exp \left (\frac{-\lb}{2a} \right )}.
\end{equation}
Here $\rho$ is the number density of the electrolyte and the degree of dissociation is assumed small. Taking the radius of a molecular ion as $a = 0.25$ nm, Eq.~\ref{eqn-rho-ions} yields an ionic concentration of $\sim$ 10$^{-13}$ mol dm$^{-3}$, for a solute concentration of 10 mM. The corresponding Debye length $\kappa^{-1} = 1/ \sqrt{4 \pi \lb \rho_{\textrm{ion}}} $ is $\sim 100 \; \mu$m.  A second distinctive feature of electrostatics in oils is the small value for the double-layer capacitance. The diffuse ion atmosphere around a charged colloid acts as a molecular condenser of capacitance, $C^{d} = \epsilon_{0} \epsilon_{r}(1+\kappa \R0)/\R0$, in the Debye-H\"{u}ckel limit \cite{4953}.  For the low ionic strengths characteristic of non-polar systems this expression reduces to $C^{d} = \epsilon_{0} \epsilon_{r}/\R0$ so that the capacitance is typically some forty times smaller in an oil than for a comparable aqueous environment. The result is that only a minute charge on a colloid in a non-polar environment is sufficient to generate an appreciable surface potential and surprisingly strong electrostatic interactions. So for
instance, the contact value of the interaction
potential (in units of $k_{B}T$) between two colloidal spheres of
radius $\R0$ and charge $e\Z0$ is, from Coulomb's law,
\begin{equation}\label{eqn-contact_pot}
    \frac{U_{0}}{k_{B}T}  =  \frac{1}{k_{B}T} \cdot \frac{\Z0^{2}e^{2}}{8\pi \epsilon_{0}
    \epsilon_{r} \R0}     =  \left ( \frac{\lambda_{B}}{2\R0} \right )
    \Z0^{2}.
\end{equation}
A 1 $\mu$m particle carrying a charge of 100 electrons, equivalent to a surface charge density of  1 $\mu$C m$^{-2}$ (about 10$^{3}$ times smaller than typical aqueous colloids), generates a very substantial electrostatic repulsion of $\approx 100$ $\kb T$ at contact. Clearly, provided colloidal particles can be efficiently charged, electrostatic interactions in a non-polar solvent will be both strong and long-ranged.

While the mechanism of charge formation in aqueous colloids is fairly well understood\cite{4953} the situation in non-polar suspensions is still far from clear\cite{4954,4436,Morrison-868,4955}.  Experiments suggest that the particle charge is a complex function of the nature of the particle surface and frequently the presence of trace amounts of water. In many of the systems studied to date, surfactants have been added to facilitate particle charging. The surfactants, which typically form reverse micelles in non-polar solvents, play an interesting dual role in these systems. First, the presence of micelles enhances the particle charge -- probably by
stabilizing countercharges in the cores of micelles. Second, micelles limits the range of the subsequent charge repulsions\cite{3771}. The vast majority of uncharged reverse micelles exist in a dynamic equilibrium with a very small fraction of positively and negatively charged micelles, generated by thermal fluctuations. This low concentration of charged micelles screens the electrostatic interactions on long length scales. Morrison, in an extensive review of the literature\cite{Morrison-868}, proposed three plausible mechanisms to account for colloid charging in non-polar surfactant systems: (A) preferential adsorption of molecular ions, surfactant aggregates or charged micelles onto the surface of a particle; (B) dissociation of surface groups with the subsequent transfer of molecular ions into the cores of reverse micelles; and (C) the adsorption of surfactant aggregates onto the particle surface, their complexation with surface groups followed by the exchange and desorption of the molecular ions into solution micelles. Much of the evidence for these mechanisms has come from electrokinetic and adsorption measurements although recently surface force measurements\cite{3305,3768} have provided direct evidence of long range electrostatic repulsions in a non-polar solvent.

The picture which has emerged to date is that the charging mechanism in non-polar environments is more subtle than that encountered in aqueous systems. A signature of this complexity is the dependence of zeta potential on surfactant concentration\cite{4437}. Focusing on colloids dispersed in low-dielectric solvents using AOT (Aerosol-OT, sodium bis(2-ethyl-1-hexyl) sulfosuccinate) several studies have reported that with increasing surfactant concentration the particle potential either monotonically decreases\cite{4908}, or more commonly display a maximum \cite{3528,3770}.  Keir et al. \cite{3770} report a highly monotonic dependence of the charge of silica in decane, which they explain qualitatively in terms of a competition between the surface binding of negative sulfosuccinate anions at low surfactant concentration and positively charged species at high [AOT] (mechanism A). Similar arguments have been invoked by McNamee et al. \cite{3768}
to account for the maximum in the interaction forces measured between two hydrophobic silica surfaces at 100 mM AOT, and by Smith et al.\cite{4908} for the gradual reduction in the zeta potential of hydrophobic TiO$_{2}$ colloids seen with increasing [AOT]. In marked contrast to these observations, Hsu et al. \cite{3771} report the striking finding that the surface potential of sterically-stabilized PMMA colloids, determined by both electrokinetic and direct interaction measurements, is \textit{independent} of AOT concentration. They propose that the different dependence of the particle charge on [AOT] is a consequence of a change in the mechanism of charging -- the polymer-coated PMMA particles charge by dissociation of surface groups (mechanism B) rather than by the adsorption of ionic species which is more frequently invoked in the case of AOT.

In this paper we re-examine the mechanism of charging of sterically-stabilized colloids in low-permittivity solvents by using the recently-developed\cite{4697} technique of single-particle optical microelectrophoresis (SPOM). An important advantage of this technique is its accuracy and sensitivity. Surface charges on the level of a few elementary charges can be reliably detected on individual colloidal particles with an uncertainty of about 0.25 $e$. To gain a broad insight into the mechanism of charging in non-polar solvents we focus on a simple model polymer-stabilized colloid with a well-defined surface chemistry and explore the particle charge produced by different species of reverse micelles. We study two surfactant and one polymeric system -- AOT, Zr(Oct)$_{2}$ [zirconyl 2-ethyl hexanoate], and the copolymer PHSA-PMMA [poly(12-hydroxystearic acid)-g-poly(methyl methacrylate)] -- each of which forms reverse micelles in dodecane. Although, our particles become negatively charged in
the presence of AOT, positive on addition of Zr(Oct)$_{2}$, and remain essentially uncharged when PHSA-PMMA is added, we find several similarities in the electrokinetics of these chemically different systems which suggests that a common physical mechanism operates in each. By combining accurate electrokinetic measurements with adsorption measurements we propose that polymer-grafted particles charge by the simultaneous adsorption of \textit{both} positively charged and negatively charged reverse micelles. Changes in the hydrophobicities of the surfactant lead to a slight excess of either positive or negative micelles on the surface and the development of a net particle charge. A statistical model of the competitive adsorption of oppositely-charged reverse micelles onto a spherical particle is analyzed and shown to be consistent with the experimental data.

\section{Experimental Section} \label{sec-exp}

\begin{table}
\begin{center}
\begin{threeparttable}[b]
\begin{tabular}{lcll}
\hline
\multicolumn{1}{c}{} & \multicolumn{1}{c}{Stabilizer} & \multicolumn{1}{c}{$\R0$ (nm)} & \multicolumn{1}{c}{$\sigma_{\R0}$} \\
\hline
RK1 & A & 42 & 0.07\tnote{a} \\
AD1 & A & 610 & 0.046\tnote{b} \\
RS1 & B & 425 & 0.10\tnote{c} \\
RS2 & B & 840 & 0.09\tnote{c} \\
RS3 & B & 1830 & 0.09\tnote{c} \\

\end{tabular}
\begin{tablenotes}
\item [a] \footnotesize{From X-ray scattering measurements.}
\item [b] \footnotesize{Static light scattering.}
\item [c] \footnotesize{Electron microscopy.}
\end{tablenotes}
\end{threeparttable}
\caption{The mean radius $\R0$ and radius polydispersity $\sigma_{\R0}$ of the colloidal PMMA particles used.} \label{table-pmma}
\end{center}
\end{table}

\subsubsection{Colloidal Particles}

Non-polar sterically stabilized poly(methyl methacrylate) (PMMA) colloids were synthesized by
a dispersion polymerization procedure, which has been described elsewhere\cite{Antl-63}. The radius of the particles was varied by adjusting the initial monomer concentration. All particles studied contained no fluorescent dyes.  Electron microscopy revealed that the particles were spherical and highly uniform in size with a mean radius $\R0$ and a radius polydispersity $\sigma_{\R0}$ (root mean square variation / mean radius) of less than $0.10$. The results are summarized in Table~\ref{table-pmma}. The particles were stabilized against aggregation by an $\sim$ 10 nm thick grafted polymer layer. The stabilizer was composed of a polymeric comb of 50 wt\% poly(12-hydroxystearic acid) (PHSA) 'teeth' and a backbone consisting of 45 wt\% PMMA and 5 wt\% poly(glycidyl methacrylate) (PGMA).  The polymeric stabilizer was covalently attached to the  particle surface. The PHSA teeth are soluble in aliphatic hydrocarbons, whilst the PMMA-PGMA backbone is insoluble so that the layer thickness is determined by the extended length of the PHSA chains. Two batches of stabilizer were used, with slightly different molecular weight distributions, as detailed in Table~\ref{table-pmma}.

\subsubsection{Micellar Solutions}

We studied three different systems of reverse micelles in decane and dodecane. Small angle neutron and X-ray scattering measurements reveal that each species forms well-defined reverse micelles at low concentrations. Literature data on the geometry and size of the reverse micelles is summarized in Table~\ref{table-micelle}. AOT (Fluka BioChemika Ultra 99 $\%$) was purified by dissolution in methanol and tumbled
with activated charcoal. The methanol was removed by rotary evaporation. The purity of the AOT was checked by a measurement of the limiting air-water surface tension. The value obtained of  27.1$\pm$ 0.1 mN m$^{-1}$ is in excellent agreement with previously reported values\cite{304}. Any increase in the water content was minimized by storing the purified surfactant in a desiccator at all times prior to use. Zirconyl 2-ethyl hexanoate (Zr(Oct)$_{2}$) was purchased from Alfa Aesar (Heysham, UK) and came as a solution in mineral spirits. The solvent was evaporated off under vacuum at 80$\degc$ and the surfactant was re-dispersed in dodecane. The polymeric PHSA-PMMA copolymer (batch A) was identical to the grafted stabilizer on the 42 nm and 610 nm PMMA colloidal particles. Analysis by GPC gave a number-average molecular weight of $M_{n} =  12550$ and a weight-average molecular weight of $M_{w} =  83400$. The PHSA-PMMA copolymer was purified by precipitation from cold methanol, dried at 45$\degc$ overnight and redissolved in dodecane at 140$\degc$.  Micellar solutions were prepared in either dodecane (Acros, 99 $\%$) or decane (Acros, 99 $\%$), which were dried with activated molecular sieves (Acros, size 4A) prior to use.

\begin{table}
\begin{center}
\begin{threeparttable}[b]
\begin{tabular}{l|lll}
 & \multicolumn{1}{c}{AOT} & \multicolumn{1}{c}{Zr(Oct)$_{2}$} & \multicolumn{1}{c}{PHSA-PMMA} \\
\hline
Geometry of reverse micelles: & \multicolumn{1}{c}{Sphere} & \multicolumn{1}{c}{Sphere} & \multicolumn{1}{c}{Cylinder\tnote{a}} \\
Hydrodynamic radius $\RM_{\textrm{h}}$ / nm & \multicolumn{1}{c}{1.6} & \multicolumn{1}{c}{1.16} & \multicolumn{1}{c}{9.17\tnote{b}} \\
Micelle volume $\VM$ / nm$^{3}$ & \multicolumn{1}{c}{17.2} & \multicolumn{1}{c}{6.5} & \multicolumn{1}{c}{2380} \\
Association number & \multicolumn{1}{c}{30} & \multicolumn{1}{c}{33} & \multicolumn{1}{c}{9} \\
Data source & \multicolumn{1}{c}{Kotlarchyk et al.\cite{3449}} & \multicolumn{1}{c}{Keir et al. \cite{3304}} & \multicolumn{1}{c}{Papworth\cite{295}} \\
\end{tabular}
\begin{tablenotes}
\item [a] \footnotesize{Small angle neutron scattering measurements \cite{295} indicate that the polymeric micelles are 28 nm in length and have a radius of 5.2 nm.}
\item [b] \footnotesize{The radius of the sphere with the same translational friction coefficient as the cylindrical micelles.}
\end{tablenotes}
\end{threeparttable}
\caption{Structural properties of reverse micelles used.}\label{table-micelle}
\end{center}
\end{table}

\subsubsection{Dispersion Formulation.}

The particle dispersions were prepared by mixing surfactant stock solutions with surfactant free particle dispersions in dried dodecane or decane. Samples were shaken vigorously before being left for 24 hours to equilibrate prior to any measurements. All solutions were sealed and stored under dry nitrogen to minimize water adsorption. The volume fraction of surfactant $\vol$ was calculated by assuming ideal mixing behavior and using the densities\cite{Bergenholtz-866,3304,295} of decane (0.73 g cm$^{-3}$), dodecane (0.75 g cm$^{-3}$), AOT (1.13 g cm$^{-3}$), Zr(Oct)$_{2}$ (2.15 g cm$^{-3}$), and PHSA-PMMA (1.04 g cm$^{-3}$).

\subsubsection{Conductivity Measurements.}

Conductivities of the micellar solutions and dispersions  were measured using a cylindrical concentric stainless steel conductivity probe (Model 627, Scientifica) at 22$\degc$. Measurements were made at an operating frequency of 15 Hz. The micellar solutions had conductivities in the range $10^{-1} \lesssim \sigma$ / pS cm$^{-1}$ $\lesssim 10^{3}$. The conductivity of the dried decane and dodecane used as solvents was recorded as $< 0.03$ pS cm$^{-1}$.   The excess ion concentration in the particle dispersions was estimated by centrifuging samples at 12000 rpm for two hours and measuring the conductivity of the upper particle-free supernatent. The viscosity of the micellar solutions was measured with a capillary Cannon Fenske viscometer operating at $25\degc$. The viscosity of dodecane at this temperature is $\eta=1.383$ mPa s$^{-1}$.

%

\subsubsection{Single Particle Optical Microelectrophoresis (SPOM).}

The electrophoretic mobility of individual colloidal particles was measured from the change in the thermal fluctuations of a particle held in an optical tweezer trap and driven by an applied sinusoidal electric field. The theory underlying the technique of single particle optical microelectrophoresis (SPOM) is discussed in detail elsewhere\cite{4697}. To perform a measurement a micropipette was used to transfer $\sim 100$ $\mu$l of a dilute suspension of particles (colloid volume fraction $\sim 3\times10^{-5}$) into an electrophoresis cell. The purpose-built cell consisted of two parallel platinum electrodes mounted in a cylindrical glass chamber and sealed with a microscope coverslip. The electrode separation was measured as 189 $\mu$m. An individual colloidal particle was optically trapped in three dimensions using the radiation pressure from a tightly-focused laser beam ($\lambda = 1064$ nm). A sinusoidal voltage with an amplitude of 5 V and a frequency of 17.5 Hz was applied. The modulation of the Brownian motion of the trapped particle produced by the applied field was measured with nanometer accuracy using an interferometric position detector. The position of the Brownian particle was collected every 10 $\mu$s for a total duration of 26 s. For each sample, data from at least 50 different individual particles was acquired, each of duration 26 s. The position detector readings were converted
into particle displacements $\Delta x(\tau)$ in the time interval $\tau$ by recording the time-dependent
mean-square voltage $\left < \Delta V^{2}(\tau) \right >$  of
five particles from the same batch of particles, with no applied field. Since the signal recorded is proportional to the displacement, $\left < \Delta V^{2}(\tau) \right >$ was fitted to the theoretical expression for the mean-squared displacement $\left < \Delta x^{2}(\tau) \right >$  of a Brownian sphere in a harmonic potential, to yield the detector calibration and the corner frequency $\omega_{c}$ of the optical trap.

We extract the electrophoretic mobility of an individual particle by calculating the spectral density $I(\Omega)$ of its Brownian fluctuations using a discrete Fourier transform. The spectrum is a sum of a Lorentzian, characteristic of Brownian motion in a harmonic potential, together with a sharp peak at the applied electric field frequency $\omega_{p}$. Integrating the spike in the power spectrum over the frequency axis yields the mean-square periodic displacement $P_{\textrm{sig}}$ of the particle. The electrophoretic mobility $\mu$ of each particle sampled was calculated from the expression\cite{4697}, $\mu^{2} E^{2} = 2P_{\textrm{sig}} (\omega_{p}^{2} + \omega_{c}^{2})$ where $E$ is the applied electric field. The sign of $\mu$ was determined by reducing the field frequency and following the oscillatory motion of the particle directly.

The electrophoretic mobilities of between 50 and 100 randomly-chosen particles were determined from each sample. The mean $\bar{\mu}$ and polydispersity $\sigma_{\mu} = \sqrt{\left< (\mu - \bar{\mu})^{2} \right >}/\bar{\mu}$ of the mobility distribution was evaluated. The mean mobility was converted into values for the mean zeta potential using the standard electrokinetic model of O'Brien and White\cite{3112}.  Numerical solutions of the coupled linearized Navier-Stokes and Poisson-Boltzmann equations were computed at each value of $\kappa \R0$, assuming that the externally applied electric field was small compared to the internal field inside the electrical double layer.

\section{Results} \label{sec-res}

\subsection{Concentration of Charged Micelles from Conductivity.} \label{sec-conc-micelles}

\begin{table}
\begin{center}
\begin{tabular}{l|lll}
& \multicolumn{1}{c}{AOT} & \multicolumn{1}{c}{Zr(Oct)$_{2}$} & \multicolumn{1}{c}{PHSA-PMMA} \\
\hline
Intrinsic viscosity, $[\eta]$ & \multicolumn{1}{c}{2.5} & \multicolumn{1}{c}{3.4} & \multicolumn{1}{c}{7.4} \\
Fraction of ionized micelles, $\CH$ & \multicolumn{1}{c}{1.5 $\times$ 10$^{-5}$} & \multicolumn{1}{c}{2.7 $\times$ 10$^{-5}$} & \multicolumn{1}{c}{3.2 $\times$ 10$^{-2}$} \\
Electrostatic charging energy, $\b \GM$ & \multicolumn{1}{c}{11.8} & \multicolumn{1}{c}{11.2} & \multicolumn{1}{c}{4.2} \\
\end{tabular}
\caption{The viscosity and conductivity of micellar solutions in dodecane.}\label{table-charge-micelle}
\end{center}
\end{table}

Because of the low dielectric constant,  reverse micelles in a solvent such as dodecane behave quite differently from charged micelles in an aqueous environment. While the total micellar charge must vanish because of electroneutrality the \textit{net} charge on each micelle fluctuates, as mobile ions are exchanged between the hydrophilic cores when micelles collide with each other. Micelle ionization is driven by spontaneous thermal fluctuations with micelle migration
in an electric field providing the main mechanism for electrical conduction in dilute
micellar solutions in oil\cite{3312,4427,4943,4942,3451}.

%

\begin{figure}[h]
\begin{center}
  \includegraphics[width=6cm]{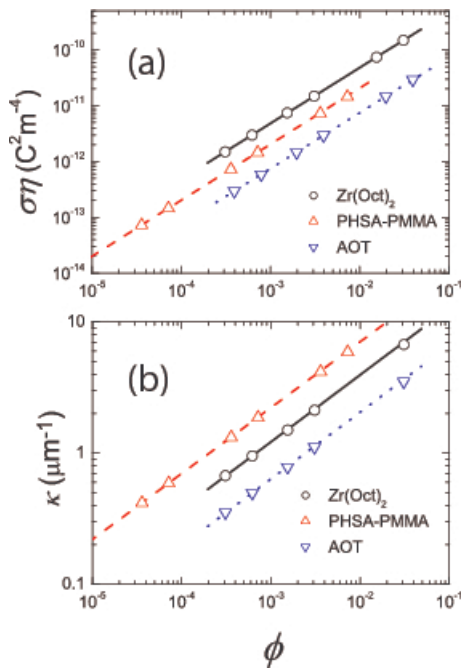}\\
  \caption{(a) $\sigma \eta$ for reverse micellar solutions in dodecane (without particles) as a function of micelle volume fraction. The lines are of unit gradient. The symbols denote measurements and the lines denote fits to Eq.~\ref{eqn_conductivity}. (b) Inverse Debye lengths determined from conductivity data.}\label{fig_visc*cond}
\end{center}
\end{figure}

The electrostatic energy  of a micelle of radius  $\RM$ carrying an excess charge $\z e$ is $\b \GMZ (\z) = \z^{2} \lb / 2\RM$, where $\b = 1 /\kb T$. If charges freely exchange between micelles
then the net charge carried by each micelle will fluctuate in time. In thermal equilibrium,
the probability $p(\z)$ of an excess charge of $\z e$ is proportional to the Boltzmann weight, $p(\z) \sim \exp (-\b \GMZ (\z))$. Since $\GMZ(\z)$ increases quadratically with $\z$ the number of multiply-charged micelles is significantly smaller than the number of singly-charged micelles. When the micelle size is much smaller than $\lb$ the concentration of multiply-charged micelles is practically negligible and may be ignored. The proportion of singly-charged micelles is fixed by the position of equilibrium in the charge exchange reaction,
\begin{equation}\label{eq_micelle-ionization}
    2\; \mathrm{uncharged \; micelles} \stequil{K} \mathrm{positive \; micelle} + \mathrm{negative \; micelle}.
\end{equation}
From the law of mass action, the equilibrium constant is $K = (n_{+} n_{-})/ n_{0}^{2}$ where $n_{+}$ and $n_{-}$ are the number densities of charged micelles and $n_{0}$ is that of uncharged micelles. Because of the practical absence of free ions, $n_{+} = n_{-}$. The fraction of ionized micelles, $\CH = (n_{+}+n_{-})/n_{0}$, is therefore $\CH = 2 \sqrt{K}$. Rewriting the constant $K$ in terms of the electrostatic energy of a singly-charged micelle $\b \GM = \lb / 2 \RM $ gives the relation,
\begin{equation}\label{eq chi}
    \CH = 2\; \ex [-\b \GM].
\end{equation}

To elucidate the nature of the charging mechanism we used three different reverse micellar systems. Conductivity measurements were used to characterize the degree of charge fluctuations in each of the solutions. For monovalent, same-sized micelles the conductivity $\sigma$ is
\begin{equation}\label{eq_conduct_def}
    \sigma = \frac{e^{2}(n_{+}+n_{-})}{\xi}
\end{equation}
where $\xi$ is the micellar friction coefficient, which depends upon the size and shape of the micelle. To discuss both spherical and cylindrical micelles, we write $\xi = 6 \pi \eta \RH$, where $\RH$ is the equivalent spherical hydrodynamic radius of the micelle. In the case of a spherical micelle $\RH = \RM$ while for a cylindrical micelle of length $l$ and diameter $d$ the equivalent radius is\cite{4433,4432}
\begin{equation}\label{eqn_friction-rod}
    \RH = \frac{l/2}{\ln p + \gamma}
\end{equation}
where $p = l/d$ is the axial ratio and $\gamma$ is an end-effect correction. Tirado and de la Torre\cite{4432}, have shown that in the range $ 2\leq p \leq 20$ relevant here, the hydrodynamics of rods are reproduced by the quadratic expression, $\gamma = 0.312 + 0.565/p  -0.1 / p^{2}$. Replacing the micelle number density by the volume fraction $\vol = n_{0} \vm$, where $\vm$ is the micelle volume, it follows immediately from Eq.~\ref{eq_conduct_def} that if micelle charging by spontaneous fluctuations is the dominant mechanism the conductivity of a dilute micellar solution should obey the simple expression,
\begin{equation}\label{eqn_conductivity}
    \sigma = \frac{e^{2}}{6 \pi \RH \eta \vm} \CH \vol.
\end{equation}
The application of this equation is complicated by the fact that the solution viscosity $\eta$ is also a function of the micelle concentration $\vol$. In the dilute regime, the relative viscosity (normalized by the solvent viscosity $\eta_{0}$) may be written in terms of the virial expansion, $\eta / \eta_{0} = 1 + [\eta] \vol + \cdots$ where quadratic and higher terms have been neglected and $[\eta]$ is the Einstein coefficient. To allow for the concentration dependence of the viscosity, capillary viscometry was used to follow the viscosity of each micellar solution.  The values obtained for $[\eta]$ are listed in Table~\ref{table-charge-micelle}. For hard spheres, the Einstein coefficient is 2.5.  Comparison with the value measured for AOT suggests that the hard sphere diameter of the AOT micelles is accurately given by Table~\ref{table-micelle}. The slightly higher Einstein coefficient observed in Zr(Oct)$_{2}$ is probably a consequence of the greater solvation of the surfactant tail layer and entrainment of solvent molecules which increases the molecular weight of the micelles and so increases $[\eta]$. The significantly larger Einstein coefficient measured for the PHSA-PMMA micelles may be accounted for at least qualitatively by the increased asymmetry of the micelles.

The charge fluctuation mechanism outlined above (Eq.~\ref{eqn_conductivity}) predicts that the product $\sigma \eta$ should depend linearly on the the micelle volume fraction $\vol$ with a gradient, $\sigma \eta / \vol$, which for fixed micelle size and shape, is purely a function of the charge fraction $\CH$. The conductivity $\sigma$ and viscosity $\eta$ of micellar solutions of AOT, Zr(Oct)$_{2}$ and the amphiphilic polymer PHSA-PMMA  were measured (with no particles) as a function of volume fraction at 22$\degc$. Figure~\ref{fig_visc*cond}(a) shows the experimentally-determined value of $\sigma \eta$ as a function of $\vol$. In each case as the concentration of micelles was increased the conductivity increased, with a linear dependence of $\sigma \eta$ upon $\vol$ being seen over two orders of magnitude change in $\vol$. From these measurements we used literature values for the size and shape of the reverse micelles formed (summarized in Table~\ref{table-micelle}) and Eq.~\ref{eqn_conductivity} to calculate the fraction $\CH$ of charged micelles. The derived values are summarized in Table~\ref{table-charge-micelle} together with the corresponding estimates of the micelle charging energy $\GM$. The value obtained for AOT, the only system where data has previously been reported, is in excellent agreement with the results of an earlier study\cite{3771}.   From the measured micelle charge fraction $\CH$, we calculate the inverse Debye length $\kappa = \sqrt{4 \pi \lb (n_{+}+n_{-})}$. The resulting values are plotted in Figure~\ref{fig_visc*cond}(b) as a function of $\vol$. Note that the electrostatic interactions between charged colloids suspended in these micellar solutions are long-ranged, with Debye lengths in the range of 0.1 $\mu$m to 10 $\mu$m.

\subsection{Surface Potentials from Electrophoretic Mobilities} \label{sec-experimental}

\begin{figure}[h]
\begin{center}
  \includegraphics[width=6cm]{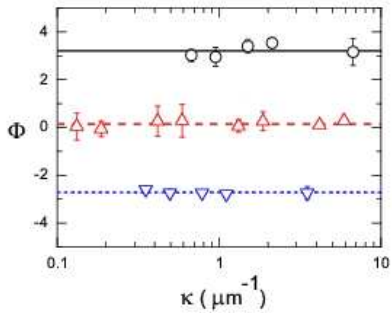}\\
  \caption{The scaled surface potential $\PH = e \bar{\phi} / \kb T$ as a function of the inverse Debye length for PMMA particles of radius $\R0 = 610 \pm 30 $ nm. The figure shows data for PMMA particles with added  Zr(Oct)$_{2}$ (circles), PHSA-PMMA (triangles), and AOT (inverted triangles). Note that for each micellar system $\PH$ is independent of the inverse Debye length $\kappa$ or equivalently the micelle concentration.}\label{fig_zeta_aotzr}
\end{center}
\end{figure}

In the absence of micelles, our particles have a very small electrophoretic mobility and are essentially uncharged. Sensitive single particle optical microelectrophoresis (SPOM) measurements on individual PMMA spheres with a radius of 610 nm gave an electrophoretic mobility of $\bar{\mu}$ = -(2.9 $\pm$ 0.2) $\times 10^{-11}$ m$^{2}$ s$^{-1}$ V$^{-1}$ and a negative zeta potential of -3.6 $\pm$ 0.4  mV \cite{4697}. However adding either 1 mM of AOT or 1.7 mM of Zr(Oct)$_{2}$ produced a dramatic change. Particles in the presence of Zr(Oct)$_{2}$ reverse micelles developed a large positive charge, and a large negative charge in the case of added AOT micelles. The electrophoretic mobilities of identically-sized particles treated with either AOT or Zr(Oct)$_{2}$ had very similar magnitudes ($\bar{\mu}_{\textrm{AOT}}$ = -(5.7 $\pm$ 0.1) $\times 10^{-10}$ m$^{2}$
s$^{-1}$ V$^{-1}$ and $\bar{\mu}_{\textrm{Zr}}$ = (6.4 $\pm$ 0.3) $\times 10^{-10}$ m$^{2}$
s$^{-1}$ V$^{-1}$) but opposite signs. As the concentration of AOT was increased from 1 mM to 100 mM and Zr(Oct)$_{2}$ from 1.7 mM to 170 mM the electrophoretic mobilities of both systems remained essentially unaltered. Addition of PHSA-PMMA solutions to our particles, at comparable levels to the AOT and Zr(Oct)$_{2}$ surfactants, gave no identifiable change in mobility ($\bar{\mu} = 3.5 \times$ 10$^{-11}$ m$^{2}$
s$^{-1}$ V$^{-1}$). The particles remained essentially uncharged. Electrophoretic mobility measurements were made at a typical electric field strength of 25 kV m$^{-1}$. High electric fields can lead to enhanced mobilities as counterions are stripped away from the particle surface, increasing the
effective charge\cite{3212}. We confirmed that this effect was unimportant in our measurements by checking that, for selected samples, $\bar{\mu}$ did not change significantly with electric field strength $|E|$, in the range $|E| < $80 kV m$^{-1}$.

Adding reverse micelles to a nonpolar suspension has two consequences. First,  it leads to particle charging and second, as discussed in Section~\ref{sec-conc-micelles}, it produces an increase in the concentration of charged micelles in solution and thus a reduction in the Debye length $\kappa^{-1}$. Using the data presented in Figure~\ref{fig_visc*cond}(b) we estimate the dimensionless inverse Debye length $\kappa \R0$ for each micelle concentration. Figure~\ref{fig_zeta_aotzr} shows the scaled particle potential $\PH = e\bar{\pot} / k_{B}T$ calculated from the mean mobility $\bar{\mu}$,  using the method of O'Brien and White \cite{3112}. While the values for $\PH$ are different in each of the three systems studied, the variation of $\PH$ with $\kappa$ is strikingly similar. In each case, we find that the surface potential is independent of the number of micelles added, over a change of 10$^{2}$ in concentration. We find $\PH = -2.72$ $\pm$ 0.07 for AOT ($0.21 \leq \kappa \R0 \leq 2.1$), $\PH = 3.2$ $\pm$ 0.2 for Zr(Oct)$_{2}$ ($0.41 \leq \kappa \R0 \leq 4.1$), and $\PH = 0.15$ $\pm$ 0.13 for PHSA-PMMA ($0.08 \leq \kappa \R0 \leq 3.6$). The equivalent particle charges are $-$(56 $\pm$ 1) $e$ [AOT], $+$(63 $\pm$ 3) $e$ [Zr(Oct)$_{2}$], and $+$(3 $\pm$ 3) $e$ [PHSA-PMMA]. This equates, for the surfactants AOT and Zr(Oct)$_{2}$, to the extremely low surface charge density of about 2 $\mu$C m$^{-2}$, some 3--4 orders of magnitude smaller than typical aqueous colloids.

\begin{figure}[h]
\begin{center}
  \includegraphics[width=6cm]{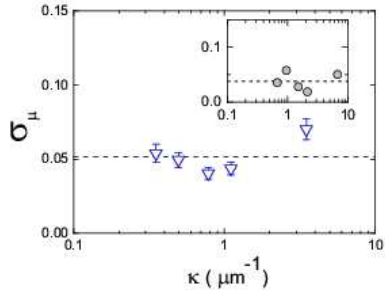}\\
  \caption{Variation of the mobility polydispersity $\sigma_{\mu}$ with the inverse Debye length $\kappa$. Conditions: PMMA particles of radius $\R0 = 610$ nm (radius polydispersity $\sigma_{R} = 0.046 \pm 0.01$) with added AOT (inverted triangles - main figure) or Zr(Oct)$_{2}$ (circles - inset). The dashed line depicts the average polydispersity: $\left < \sigma_{\mu} \right > = 0.052 \pm 0.01$ (main - AOT); $\left < \sigma_{\mu} \right > = 0.038 \pm 0.02$ (inset - Zr(Oct)$_{2}$). }\label{fig_poly_aotzr}
\end{center}
\end{figure}

Despite the micelles of AOT, Zr(Oct)$_{2}$ and PHSA-PMMA being chemically different we see several similar features. A key result of our study is that the surface potential remains essentially independent of the number of micelles added. This observation agrees with the results of more limited experiments on the AOT/PMMA system reported by Hsu et al. \cite{3771}, although this study found a significantly higher surface potential $\PH = -5.5$ $\pm$ 0.2 than the values obtained here. The negligible effect on $\PH$ of the number of micelles in solution is inconsistent with charging arising from the dissociation of surface groups. If ionization of surface groups was the source of the particle charge then increasing micelle concentration should increase the concentration of counterions in the continuous phase, and lead  to an increase in the charge per particle. Consequently the experimental observations
rule out surface dissociation as a charging mechanism. However the lack of dependence of  $\PH$ on micelle concentration is also difficult to explain if we assume charging is caused by \textit{simple} ion adsorption. The tiny charge densities observed means that the particle surface is far from saturated. Consequently, increasing the amount of adsorbable charged species in solution, by adding more micelles, should lead to more ion adsorption and hence an increased particle charge. The prediction is inconsistent with our observations and apparently also rules out simple adsorption as a charging mechanism. In Section~\ref{sec-model} we discuss an alternative mechanism of charging which is both physically acceptable and compatible with our experimental observations.

Single particle optical microelectrophoresis (SPOM) measurements were performed on 50 -- 100 individual colloidal particles at each micelle concentration, for each of the three micellar systems studied. The raw data was thus a scatter plot of mobility with one point from each individual particle. From this data a mobility distribution $P(\mu)$ was determined. In all cases this distribution was well fitted by a Gaussian, characterized by a mean mobility and a
polydispersity $\sigma_{\mu}$ defined by
\begin{eqnarray}
  \bar{\mu} &=& \int_{0}^{\infty} P(\mu) \mu \rm{d}\mu  \\
  \sigma_{\mu} &=& \frac{1}{\mid \bar{\mu} \mid}
  \left ( \int_{0}^{\infty} P(\mu) (\mu- \bar{\mu})^{2}
  \right )^{1/2}.
\end{eqnarray}
Figure~\ref{fig_poly_aotzr} displays the variation of $\sigma_{\mu}$ with the inverse Debye length following the addition of AOT to a suspension of 610 nm PMMA particles. The mobility distribution is surprisingly narrow with a width of order 5\%. As with other electrokinetic parameters, there is no systematic variation with micelle concentration. Averaging the measured values for the mobility polydispersity together gave $\left < \sigma_{\mu} \right > = 0.052 \pm 0.01$. Repeating the procedure for the Zr(Oct)$_{2}$ surfactant (data shown in inset of Figure~\ref{fig_poly_aotzr}) gave a very similar value, $\left < \sigma_{\mu} \right > = 0.038 \pm 0.02$. These two results correspond rather nicely to the value for the size polydispersity $\left < \sigma_{\R0} \right > = 0.046 \pm 0.01$ measured for the 610 nm particles.

\begin{figure}[h]
\begin{center}
  \includegraphics[width=6cm]{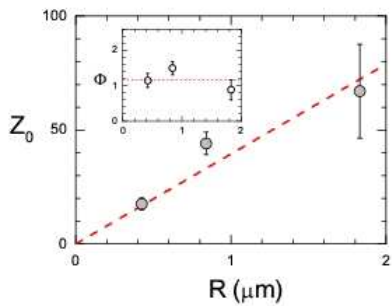}\\
  \caption{Variation of the particle charge $\Z0$, in units of the elementary charge, with the radius $\R0$. The PMMA particles were prepared using stabilizer B and the dispersions contained 100 mM of AOT.  The dashed line depicts the
linear relationship, $\Z0 = A \R0 / \lb$, with $A = 1.12$. The size dependence of the corresponding scaled surface potential $\PH = e \phi / \kb T$, evaluated from the electrokinetic model of O'Brien and White, is plotted in the inset figure. The data is consistent with a size-independent surface potential $\left < \PH \right > = 1.18 \pm 0.18$. }\label{fig_phi_radius}
\end{center}
\end{figure}

To understand the significance of this agreement we restate a few of the key results of standard electrokinetic theory. In the H\"{u}ckel limit appropriate here (small $\kappa \R0$) the scaled electrophoretic mobility, defined as
\begin{equation}\label{eqn-scaled-mu}
    \tilde{\mu} = \frac{3}{2} \frac{\eta e}{\epsilon_{0} \epsilon_{r} \kb T} \mu
\end{equation}
assumes the value $\tilde{\mu} = \Z0 \lb / \R0 = \PH$ where $\Z0$ is the effective charge on the particle and $\PH$ is the reduced surface potential. The relationship between $\mu$ and $\Z0$ is linear so the mobility polydispersity and the charge polydispersity should be equal. The numerical correspondence seen between $\left < \sigma_{\mu} \right >$ and $\left < \sigma_{\R0} \right >$ therefore translates into a linear dependence of the electrophoretic effective charge on the particle radius. We have checked this correlation for a wider range of radii by preparing a number of differently-sized particles and measuring their mobilities in the presence of a fixed concentration of AOT (100 mM). Figure~\ref{fig_phi_radius} shows the resulting variation in the measured electrophoretic charge $\Z0$ with the particle radius. The data is consistent with a linear dependence of the charge on the radius given the errors in $\Z0$ and when taken together with the
correspondence between $\sigma_{\mu}$ and $\sigma_{\R0}$, evident in Figure~\ref{fig_poly_aotzr}, strongly supports the case for a linear correlation between $\Z0$ and $\R0$. We find $\Z0 = A \R0 / \lb$ with $A = - 1.1 \pm 0.1$ for the samples in Figure~\ref{fig_phi_radius} and $A \approx - 2.6$ for the 610 nm particles with added AOT included in Figure~\ref{fig_zeta_aotzr}. The difference in the coefficients $A$ determined for different batches of particles is probably a consequence of the different stabilisers used in their synthesis (see Table~\ref{table-pmma}).

A similar linear dependence of the effective charge on the particle radius has been observed previously for both aqueous \cite{4230} and non-aqueous systems \cite{4438}. Garbow et al. \cite{4230} found $\Z0 = A \R0 / \lb$ with a coefficient $A \approx 2$ for very dilute aqueous suspensions of highly charged poly(styrene) spheres under near salt-free conditions. While Strubbe et al.\cite{4438}  found $A \approx 1$ in a dodecane suspension of pigment particles with added poly(isobutylene)succinimide. A linear relationship between particle charge and radius has been predicted theoretically \cite{3218} and observed in computer simulations \cite{3206} but only in the charge saturation limit where electrostatic interactions dominate. However the predicted value for the linear coefficient ($A \approx 10$) is significantly larger than the values recorded here. On the basis of this discrepancy, the low values recorded for the surface potentials, and the observation that the coefficient $A$ varies with the nature of the particle surface, we conclude that the linear dependence of $\Z0$ upon the radius $\R0$ seen in Fig.~\ref{fig_phi_radius} can not be explained in terms of charge saturation.

\subsection{Adsorption of Surfactant} \label{sec-micelle-adsorption}
%

In recent years, the structures of amphiphiles at solid surfaces has been extensively studied using techniques such as neutron reflection, fluorescence spectroscopy and atomic force microscopy. The molecular organization seen is surprisingly complex. A variety of structures have been proposed, ranging from spherical aggregates resembling bulk micelles, through cylinders and perforated layers, to uniform continuous layers \cite{5028}.
While the self-assembly of surfactants on polar surfaces from aqueous solution has been extensively studied, a lot less attention has been paid to the adsorption of surfactants from organic solvents and the information, when available, is limited. For the specific case of the anionic surfactant, AOT, fluorescence studies \cite{4934} have revealed the presence of reverse ``micelle-like'' surfactant aggregates for adsorption onto hydrophobic graphite particles, from cyclohexane.
The adsorption isotherm of the surfactant AOT on PMMA particles has been measured by Kitahara et al. \cite{4437}. The adsorption increases sharply at low concentrations suggesting a high affinity of the surfactant for the surface of the particle before reaching a plateau value at high concentrations.

\begin{figure}[h]
\begin{center}
  \includegraphics[width=6cm]{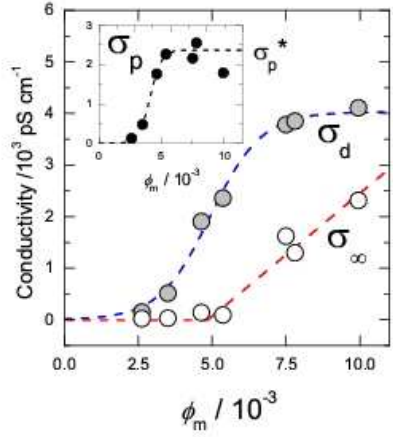}\\
  \caption{The conductivity of a suspension of small PMMA particles (RK1, $\R0 = 42$ m,) as a function of the volume fraction $\vol$ of swollen AOT reverse micelles in decane. The colloid volume fraction was fixed at $\volc = 0.08$. Filled circles: conductivity of suspension $\sigma_{d}$; open circles: conductivity of supernatent $\sigma_{\infty}$; filled circles (inset): contribution to suspension conductivity from particles alone, $\sigma_{p} = \sigma_{d} - \sigma_{\infty}$. Note, for $\vol \leq 5 \times 10^{-3}$ the conductivity of the suspension is finite, although no micelles are detectable in the supernatent ($\sigma_{\infty} \sim 0$). This indicates that particles charge by the adsorption of charged micelles.}\label{fig-conductivity-suspension}
\end{center}
\end{figure}

If surfactant micelles are adsorbed onto the surface of the particle then the number of free micelles in solution must decrease. Accordingly we expect, from the charge fluctuation model, the number of charged micelles and hence the solution conductivity to decrease. To confirm this, we measured the conductivity of a dispersion of 42 nm particles with a constant colloid volume fraction, $\volc = 0.08$,  suspended in a decane solution of AOT micelles. To amplify the conductivity changes we increased the size of the reverse micelles by adding water, keeping the molar ratio $w$ of water to AOT fixed at 40.8 so that the micelle radius is fixed. Small angle neutron scattering measurements \cite{Bergenholtz-866} indicate that, under these conditions, the reverse micelles have a radius of $\RM = 7.8$ nm, independent of concentration. The conductivity of the resulting dispersion, $\sigma_{d}$, has contributions from (1) the motion of the charged particles and their accompanying diffuse layer of micellar counterions ($\sigma_{p}$), and (2)
excess micellar ions ($\sigma_{\infty}$), so that $\sigma_{d}= \sigma_{p} + \sigma_{\infty}$. To distinguish these terms, we used centrifugation to separate the colloidal particles from excess micellar ions. The conductivity
of the supernatent provides an estimate for $\sigma_{\infty}$, since the neutral particle sediment contains only the charged particles and associated counterions. Figure~\ref{fig-conductivity-suspension} shows the dependence of $\sigma_{d}$ and $\sigma_{\infty}$ on the total volume fraction $\vol$ of reverse micelles added to the system.
In the dispersion, at low micelle concentrations ($\vol \leq 5 \times 10^{-3}$) there are essentially no charged micelles left in solution and the conductivity of the supernatent is practically zero. For the same concentrations Fig.~\ref{fig-conductivity-suspension} reveals that the particles become increasingly highly charged as $\sigma_{d}$ rises rapidly with $\vol$. The contribution to the suspension conductivity from the particles alone,
 $\sigma_{p}$, is plotted in Figure~\ref{fig-conductivity-suspension}(b). Clearly as micelles are added to the dispersion, $\sigma_{p}$ increases before finally reaching a plateau of $\sigma_{p}^{*} \approx ( 2.4 \pm 0.2) \times 10^{3}$ pS cm$^{-1}$ at high micelle concentrations. The dependence of the particle
conductivity on the micellar volume fraction, $\vol$, is reminiscent of a Langmuir isotherm with the rapid rise at low concentrations suggesting a high affinity interaction between the particle surface and micelles.

The value of the plateau conductivity $\sigma_{p}^{*}$ provides an estimate of the particle charge. Since the double layer is extremely diffuse ($\kappa \R0 \ll 1$), the charged particle and counterions move independently of each other in an applied electric field. The particle conductivity $\sigma_{p}^{*}$ may therefore be expressed as
\begin{eqnarray} \label{eqn-sigma-p}
  \sigma_{p}^{*} &=& \frac{\Z0^{2}e^{2} }{6 \pi \eta \R0} n_{p} +  \frac{|\Z0| e^{2} }{6 \pi \eta \RM}n_{p} \\
   &=&  \frac{e^{2} \volc}{8 \pi^{2} \eta \R0^{3} \RM} \left [ \left(\frac{\RM}{\R0} \right) \Z0^{2} + |\Z0| \right ] \nonumber
\end{eqnarray}
where $n_{p} = 3 \volc / 4 \pi \R0^{3}$ is the particle number density. Here the first term on the right-hand side of Eq.~\ref{eqn-sigma-p} arises from the motion of the particles, and the second term is due to counterions. Using this expression and the measured value of $\sigma_{p}^{*}$ we estimate the mean charge of each particle as 4.1 $e$. While the particles used for the conductivity experiments are too small for measurements of the electrophoretic mobility, the equivalent dimensionless charge $\Z0 \lb / \R0$ is 2.8, in reasonable agreement with the value found for the larger 610 nm polymer particles by SPOM ($\Z0 \lb / \R0 = 2.58 \pm 0.04$).
%

\section{Charging mechanism} \label{sec-model}

\begin{figure}[h]
\begin{center}
  \includegraphics[width=6cm]{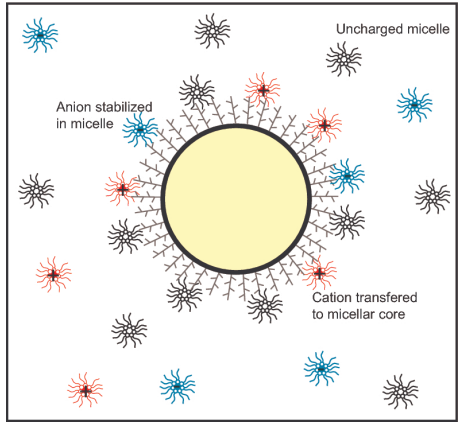}
  \caption{A micelle-decorated polymer particle. The decorated particle
acquires a surface charge by the adsorption of an excess of either positively or negatively charged micelles.}\label{fig_ch_mech_3}
\end{center}
\end{figure}
%

In this section we analyze a charge regulation model of the micelle-decorated colloid, illustrated in Figure~\ref{fig_ch_mech_3}. We assume that oppositely-charged micelles compete with each other and with uncharged micelles for the same binding sites on the surface of the particle. Using equilibrium statistical mechanics we examine the partitioning of positive and negatively-charged micelles between solution and the particle surface and show that the model is consistent with the data presented in Section~\ref{sec-res}. The model is essentially a variant of the classical charge regulation model of amphoteric surfaces first introduced by Chan et al. \cite{4896} and applied recently to particle charging by Strubbe and coworkers\cite{4881}. Below we present a simple physical derivation of the charge regulation model which is tailored to the specific problem under consideration. For a more detailed general treatment the reader is referred to the original paper\cite{4896}.
%
%
%

\subsection{Competitive Adsorption of Charged Micelles.}\label{section-competitive}

A non-polar surfactant solution contains a random mixture of  neutral and charged micelles. Reverse micelles frequently display a short-range attraction in organic media due to a mutual interpenetration of surfactant tails\cite{4987} or a solvent-mediated depletion interaction\cite{4966,4986}. Similar attractive interactions probably operate between micelles and the hydrophobic polymer chains which coat a sterically-stabilized colloid. As a result, we expect the surface of a particle to be decorated with a random mixture of charged and uncharged micelles. In our model the number of micelles that can adsorb is limited by $N$, the number of available surface sites per particle. The extent of absorption is controlled by $M$, the number of free micelles per particle. The net charge on the particle will fluctuate in time as the number of positively-charged and negatively-charged micelles adsorbed change as a result of exchange and/or  charging reactions. We assume that the partitioning of micelles between surface and solution is determined purely by equilibrium energetics -- the differences in free energy between adsorbed and free micelles of either positive $\GP$, negative $\GN$ or neutral charge $\GU$. Without loss of generality we consider the situation where $\GP < \GN$ so that positive micelles are more strongly adsorbed than negative micelles and the particle develops a net positive surface potential $\pot$. The particle charge becomes progressively more positive as more
positive micelles adsorb. However an electrostatic feedback limits the maximum charge. As micelles adsorb, the electrostatic repulsions between the particle and free positively-charged micelles become increasingly dominant. At some point, approaching positive micelles are repulsed and micelles with the opposite sign are attracted to the particle surface. The charge on the particle is accordingly regulated by the competition of the different micellar species with each other for the available surface sites.

To analysis this situation, we focus on the energetics of adsorption of a single positive micelle onto a colloidal particle. The chemical potential difference $\DMU$ between adsorbed and free micelles has three contributions. First, there is the loss of the translational free energy when a micelle is bound to the particle surface. Off setting this energetic cost is the gain in surface free energy as the micelle is adsorbed to any one of a large number of vacant surface sites. Finally there are the energetic terms: the energy of adsorption, $\GP$ and the electrostatic energy $e \phi$ arising from the Coulomb repulsion between the charged micelle and a particle with potential $\phi$.

The translational entropy of the positively-charged micelles after $\ZP$ micelles have been absorbed on the surface of a particle is
\begin{equation}\label{eq_trans entropy}
-S_{\mathrm{T}}=\kb\left[\left(\frac{\MP \CH
}{2}-\ZP\right)\ln \left(\frac{\MP \CH}{2}-\ZP\right)-\left(\frac{\MP \CH}{2}-\ZP\right)\right],
\end{equation}
where $\MP \CH /2-\ZP$ is the remaining number of positive micelles in solution. The change in translational free energy per unit positive micelle is accordingly
\begin{equation}\label{eq trans mu}
    -\b T \mathrm{d}S_{\mathrm{T}} /  \mathrm{d} \ZP = - \ln (\MP \CH /2-\ZP).
\end{equation}

To estimate the configurational entropy of the surface phase of positive micelles we suppose there are $\ZP$ ions and $\NP- \ZN - \ZU$ unoccupied sites. The total number of arrangements of the surface phase is
\begin{equation}\label{eq_config}
\Omega = \genfrac{(}{)}{0pt}{}{\NP-\ZN-\ZU}{\ZP}
=\frac{(\NP-\ZN-\ZU)!}{(\NP-\Sigma)!\ZP!}.
\end{equation}
where $\Sigma$ is the total number of micelles adsorbed, $\Sigma = \ZP + \ZN + \Z0$. If we assume that all of these arrangements are equally probable then the configurational entropy will include a term $S_{\mathrm{C}} = \kb \ln \Omega$, which using Stirling's approximation reduces to
\begin{equation}\label{eq config entropy}
- \kb \left [ \ZP \ln \ZP / (\NP-\ZN-\ZU) + (\NP-\Sigma) \ln  (\NP-\Sigma) / (\NP-\ZN-\ZU) \right ].
\end{equation}
Differentiation with respect to $\ZP$ gives the surface entropy contribution to the free energy change per unit micelle as
\begin{equation}\label{eq surface mu}
    -\b T \mathrm{d}S_{\mathrm{C}} /  \mathrm{d} \ZP = - \ln (\NP-\Sigma) / \ZP.
\end{equation}

The equilibrium concentration of positive micelles is determined by the condition that the chemical potential of the charged micelles is the same at the surface as in solution. Combining the entropic (Eqs.~\ref{eq trans mu} and \ref{eq surface mu}) and energetic contributions gives the difference in chemical potential between bound and free positive micelles as,
\begin{equation}\label{eq delta mu}
    \b \DMU = - \ln (\NP-\Sigma) / \ZP - \ln (\MP \CH /2-\ZP) + \b \GP + \PH,
\end{equation}
where $\PH= e \phi / \kb T$ is the dimensionless particle potential. In equilibrium $\DMU = 0$. To simplify the equations from here on we assume that the colloid concentration is sufficiently small that the number of free micelles exceeds the number adsorbed so that $\CH \MP /2 \gg \ZP$. This is a reasonable assumption in the case of the single particle data presented in Section~\ref{sec-res}. In this regime, Eq.~\ref{eq delta mu} rearranges to an expression for the number of positive micelles adsorbed,
\begin{equation}\label{eq ZP}
    \ZP = \MP (\NP-\Sigma) \KP \exp (-\PH)
\end{equation}
where the equilibrium constant $\KP = \exp \left [ - \b (\GP + \GM) \right ]$ is independent of $\NP$ and $\MP$. Note that $\KP$ refers to the two-stage process; adsorption of an uncharged micelle onto the particle surface followed by ionization of the adsorbed micelle. Typically $\GP + \GM > 0$ so the free energy of the adsorbed charge micelle is higher than the free micelle.   As expected, Eq.~\ref{eq ZP} reveals that $\ZP$ is a sensitive function of the surface potential, decreasing as $\PH$ becomes more positive due to the increased electrostatic repulsion between the positive micelle and positive surface.

Applying similar arguments to the adsorption of the negative and uncharged micelles yields expressions for $\ZN$ and $\ZU$
\begin{eqnarray}\label{eq ZN and Z0}
  \ZN  &=& \MP (\NP-\Sigma) \KN  \exp (\PH) \nonumber \\
  \ZU &=& \MP (\NP-\Sigma) \KU
\end{eqnarray}
with the corresponding equilibria constants
\begin{eqnarray}\label{eq equil const}
  \KN &=& \exp \left [ - \b (\GN + \GM) \right ] \nonumber \\
  \KU &=& \exp \left [ - \b \GU \right ].
\end{eqnarray}

The number of charged micelles adsorbed depends upon the potential and is therefore unknown. Rather than specify the potential  we determine $\PH$ self-consistently as follows: The equilibrium particle charge $\Z0$ is determined by the difference in the number of positive and negative micelles adsorbed upon the surface
\begin{equation}\label{eq Z0}
    \Z0 = \ZP - \ZN.
\end{equation}
The relation between $\PH$ and $\Z0$ is found by solving the Poisson-Boltzmann equation for a spherical particle. In the Debye-H\"{u}ckel limit where the surface potential is small compared with the thermal energy, $\PH \ll 1$, the Poisson-Boltzmann equation may be linearized and solved analytically with the well known result
\begin{equation}\label{eq pot charge}
    \PH = \frac{\Z0 \lb}{\R0 (1 + \kappa \R0)}.
\end{equation}
Substitution of Eq.~\ref{eq Z0} into  Eq.~\ref{eq pot charge} yields an explicit expression for $\PH$ in terms of the equilibria constants $\KP$, $\KN$ and $\KU$.

\subsubsection{Analytical solution for low surface coverage.} \label{sec-low-coverage}

In many instances the number of sites for adsorption is significantly larger than the total number of micelles adsorbed ($\NP \gg \Sigma$) so the particle surface is only sparsely covered with micelles. In this regime, the calculation of the equilibrium surface potential $\PH$ is considerably simplified and the system of equations (\ref{eq ZP})-- (\ref{eq pot charge}) can be solved analytically.

\begin{figure}[h]
\begin{center}
  \includegraphics[width=6cm]{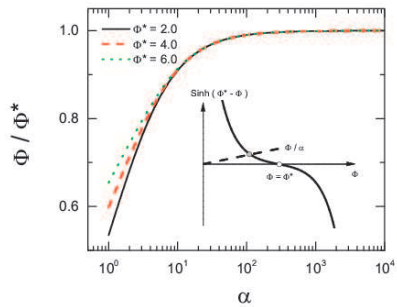}\\
  \caption{Typical solutions of Eq.~\ref{eq phi transdent} for the equilibrium surface potential $\PH$, scaled by the plateau potential $\Ps$, as a function of the charging coefficient $\alp$. Note that $\PH$ is independent of $\alp$ when $\alp \gtrsim 10^{2}$. The inset diagram gives a graphical depiction of the solution of
  Eq.~\ref{eq phi transdent}.}\label{fig_phi large N}
\end{center}
\end{figure}

Substituting Eqs.~\ref{eq ZP} and \ref{eq ZN and Z0} into \ref{eq pot charge} yields an approximate expression for the surface potential $\PH$,
\begin{equation}\label{eq phi}
    \PH = \frac{\lb}{\R0 (1 + \kappa \R0)} \MP \NP \left [ \KP \exp (-\PH) - \KN \exp (\PH) \right ].
\end{equation}
The limit $\MP \NP \gg 1$ proves to be a useful guide to understanding and solving the adsorption behavior. To simplify Eq.~\ref{eq phi} we introduce the saturation potential,
\begin{equation}\label{eq phi plateau}
    \Ps = \frac{\b}{2} ( \GN - \GP).
\end{equation}
Eq.~\ref{eq phi} may then be rewritten as the transcendental equation
\begin{equation}\label{eq phi transdent}
    \PH = \alpha \sinh (\Ps -\PH)
\end{equation}
where the charging coefficient,
\begin{eqnarray}\label{eq alpha}
    \alp & = &  \frac{2\lb}{\R0 (1 + \kappa \R0)} \MP \NP \sqrt{\KP \KN} \nonumber \\
    & = & \frac{2\lb}{\R0 (1 + \kappa \R0)} \sqrt{\ZP \ZN}
\end{eqnarray}
is $\PH$-independent. The second expression for $\alp$ in Eq.~\ref{eq alpha} follows from Eqs.~\ref{eq ZP} and \ref{eq ZN and Z0} in the limit where $\NP \gg \Sigma$. The solution to Eq.~\ref{eq phi transdent} may be found by a simple graphical construction, illustrated in the inset diagram of Fig.~\ref{fig_phi large N}. The equilibrium potential $\PH$ is defined by the intersection between the straight line $y = \PH / \alp$ and the curve $y = \sinh (\Ps -\PH)$. A moment's reflection shows  that, as the charging coefficient $\alp$ increases, the equilibrium surface potential $\PH$  approaches asymptotically the saturated surface potential $\Ps$.

The saturation of the equilibrium surface potential with increasing $\alp$ is confirmed by a full numerical solution of Eq.~\ref{eq phi transdent} -- typical examples of which are depicted in Figure~\ref{fig_phi large N}. The equilibrium potential first increases monotonically with increasing $\alp$ before finally leveling off at $\PH = \Ps$. In the plateau regime where $\alp \gtrsim 10^{2}$, the predictions of the adsorption model are especially simple.  Here the surface potential is a constant, independent of the radius of the particle $\R0$, the micelle concentration $\MP$, the number of surfaces sites $\NP$ or the Debye length $\kappa^{-1}$ (see Figure~\ref{fig_phi large N}). Indeed Eq.~\ref{eq phi plateau} reveals that the limiting  potential $\Ps$ is a function \textit{only} of the relative adsorption strengths of the surface for charged micelles and is totally unaffected by the composition of the system. Similar conclusions follow from the more general model of particle charging introduced by Strubbe et al.\cite{4881}

\begin{figure}[h]
\begin{center}
  \includegraphics[width=6cm]{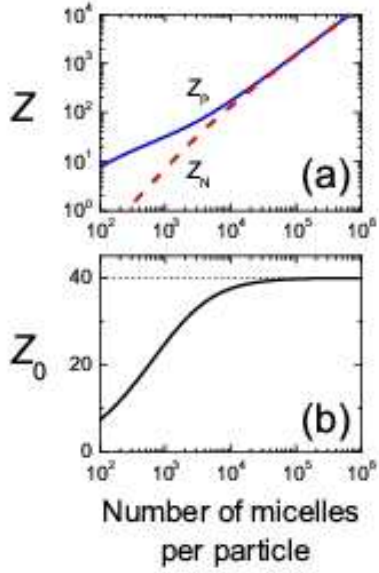}\\
  \caption {Competitive adsorption of charged micelles.  Variation of (a) the average number of bound positive ($\ZP$) and negative ($\ZN$) micelles and, (b) the equilibrium particle charge ($\Z0$) with the number of free micelles per particle. The curves correspond to the numerical solution of Eqs.~\ref{eq ZP}, \ref{eq ZN and Z0} and \ref{eq phi} and assume low surface coverage, $\NP \gg \ZP + \ZN + \Z0$. The parameters $\R0 = 20 \lb$, $\NP = 10^{6}$, $\kappa \R0 = 0$, $\ln \KP = -16$, and $\ln \KN = -20$ were used. The dashed line in (b) denotes the saturation charge $\Zs = \R0 \Ps / \lb$.}\label{fig_Zp_Zn}
\end{center}
\end{figure}

The independence of the equilibrium particle charge on the concentration of micelles is illustrated by the calculations depicted in Figure~\ref{fig_Zp_Zn}. A fixed value of $\R0 = 20 \lb$ was chosen for the particle radius. The equilibrium particle charge $\Z0$ and the number of adsorbed positive ($\ZP$) and negative micelles ($\ZN$) were calculated from Eqs.~\ref{eq ZP}, \ref{eq ZN and Z0}, and \ref{eq phi transdent} for solution micelle concentrations between $\MP =10^{2}$ and 10$^{6}$. Free energies of micelle adsorption and ionization were fixed at $\b (\GP + \GM) = 16$ and $\b (\GN + \GM) = 20$ so that the particle charges positive. Figure~\ref{fig_Zp_Zn}(a) reveals that the number of positive and negatively-charged micelles adsorbed increases uniformly with micelle concentration and is essentially linear with $\MP$ at high concentrations. The equilibrium particle charge $\Z0$, which is equal to $\ZP-\ZN$, however shows a very different dependence on $\MP$ as a result of a electrostatic feedback mechanism which limits the growth of the equilibrium charge. Figure~\ref{fig_Zp_Zn}(b) shows that the particle charge $\Z0$ is initially proportional to $\MP$ for low micelle concentration before finally saturating at high micelle concentrations to a value $\Zs = \R0 \Ps / \lb$. Any further increase in $\MP$ causes an additional \textit{equal} number of positive and negative micelles to be adsorbed onto the particle. Accordingly the population difference $\ZP - \ZN$ does not vary with $\MP$ and so the particle charge remains constant. The saturated particle charge $\Zs$ is immune to any change in $\MP$ provided there are still empty surface sites on the particle.

\begin{figure}[h]
\begin{center}
  \includegraphics[width=6cm]{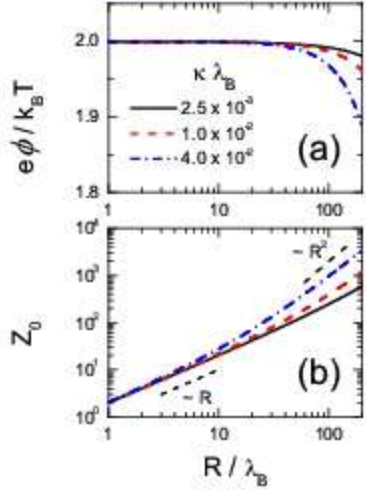}\\
  \caption
  {Competitive adsorption of charged micelles.  Predicted variation of (a) the scaled surface potential $e \pot / \kb T$ and, (b) the equilibrium particle charge $\Z0$ with the particle radius $\R0 / \lb$. Each curve corresponds to calculations at a fixed inverse Debye length $\kappa \lb$ (values given in legend). Eqs.~\ref{eq ZP}, \ref{eq ZN and Z0} and \ref{eq phi} were solved at low surface coverage for charging parameters: $\MP = 10^{6}$, $\NP = 10^{6}$, $\ln \KP = -16$, and $\ln \KN = -20$.}\label{fig_vary R}
\end{center}
\end{figure}

Figure~\ref{fig_vary R} shows the size dependency of the charging mechanism. The equilibrium surface potential $\PH$ is again seen to be practically independent of the particle radius $\R0$ upto the largest particle sizes where $\PH$ shows a relatively small decrease. The drop in $\PH$ is more marked in strongly screened systems where the charging coefficient $\alpha$ (Eq.~\ref{eq alpha}) is reduced. The particle charge is linear with increasing radius in the regime where $\kappa \R0 \lesssim 1$ before becoming quadratic in the strongly-screened limit where $\kappa \R0 \gtrsim 1$.

\subsubsection{Numerical solution for high surface coverage.} \label{sec-high-coverage}

When the number of micelles adsorbed is comparable to the number of available surfaces sites then the calculation of the equilibrium potential $\PH$ is more involved. In this section we outline the general solution and discuss the implications of a finite surface coverage for the charging mechanism.

The surface populations of micelles is given by
\begin{eqnarray}\label{eq_ZP_ZN_Z0}
  \ZP &=& \MP (\NP-\Sigma) \KP  \exp (-\PH) \nonumber \\
  \ZN  &=& \MP (\NP-\Sigma) \KN  \exp (\PH) \nonumber \\
  \ZU &=& \MP (\NP-\Sigma) \KU.
\end{eqnarray}
Substituting Eq.~\ref{eq_ZP_ZN_Z0} into the definition of $\Sigma$ and rearranging produces an explicit expression for $\Sigma$ in terms of the unknown surface potential $\PH$,
\begin{equation}\label{eq_sigma}
    \Sigma = \frac{\MP \NP \left [ \KP  \exp (-\PH) + \KN  \exp (\PH) + \KU \right ]}{1+ \MP \left ( \KP  \exp (-\PH) + \KN  \exp (\PH) + \KU \right )}.
\end{equation}
The surface potential is determined self-consistently from the particle charge using the Debye-H\"{u}ckel expression,
\begin{equation}\label{eq_pot_DH}
     \PH = \frac{\lb (\ZP - \ZN) }{\R0 (1 + \kappa \R0)}.
\end{equation}
Substitution of Eq.~\ref{eq_ZP_ZN_Z0} into Eq.~\ref{eq_pot_DH} yields a nonlinear expression for $\PH$ valid at all surface coverages,
\begin{equation}\label{eq_general_eqn_phi}
        \PH = \frac{\lb}{\R0 (1 + \kappa \R0)} \MP (\NP - \Sigma) \left [ \KP \exp (-\PH) - \KN \exp (\PH) \right ]
\end{equation}
where Eq.~\ref{eq_sigma} gives $\Sigma$ in terms of $\PH$. In the limit of low coverage, $\NP \gg \Sigma$, equation~\ref{eq_general_eqn_phi} reduces to Eq.~\ref{eq phi}.

\begin{figure}[h]
\begin{center}
  \includegraphics[width=6cm]{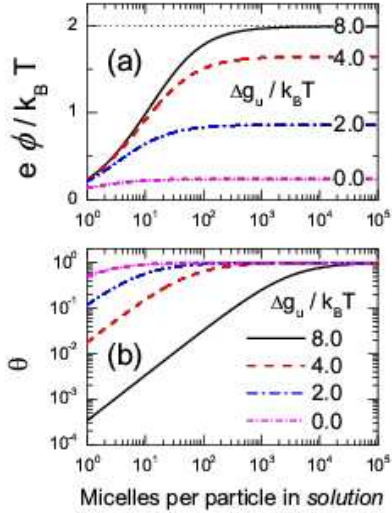}\\
  \caption
  {The dependence of (a) the equilibrium surface potential, and (b) the fraction of surface sites occupied by micelles, upon the number of micelles per particle in solution. The curves are plotted  for a number of different values for the uncharged micelle adsorption energy $\GU$. The dotted line in (a) is the saturated potential $\Ps$ according to Eq.~\ref{eq phi plateau}. The results were computed from Eqs.~\ref{eq_sigma}  and \ref{eq_general_eqn_phi} with $\R0 = 20 \lb$, $\NP = 10^{6}$, $\ln \KP = -12$, $\ln \KN = -16$, and $\kappa \R0 = 0$. }\label{fig_full_calc}
\end{center}
\end{figure}

The effect on the surface potential of high degrees of micelle adsorption  is illustrated in Figure~\ref{fig_full_calc}. The plots were computed from Eq.~\ref{eq_general_eqn_phi} for a particle of radius $\R0 = 20 \lb$ with $\NP = 10^{6}$, $\b (\GP + \GM) = 12$, and  $\b (\GN + \GM) = 16$ in the regime $\kappa \R0 \ll 1$. The fraction $\Theta = \Sigma / \NP$ of the particle's surface covered by adsorbed micelles was varied by adjusting the adsorption energy $\GU$ of uncharged micelles. If the uncharged micelles are only weakly adsorbed (for example, $\b \GU = 8$) the surface is completely covered only at high micelle numbers, $\MP \geq 10^{5}$. In this case the
surface potential $\PH$ displays a similar dependence on $\MP$ as that found previously (Figure~\ref{fig_phi large N}) and $\PH$ reaches the asymptotic limit, $\PH = \Ps$. Increasing the number of uncharged micelles adsorbed has two consequences: (I) The surface potential saturates at a lower value ($\PH < \Ps$) which decreases with increasing micelle adsorption and, (II) the concentration of micelles at which the potential saturates shifts to smaller $\MP$, as all absorption sites on the particle surface become occupied.

\subsection{Analysis of experimental data}

The analysis above demonstrates that the main physical quantity which controls the particle charge  is the surface coverage $\Theta$. We now proceed to estimate $\Theta$ for our experimental system. If the reverse micelles  pack in a triangular tesselation on the surface of the colloid  then the maximum number of micelles that can be accommodated is $\NP = (2 \pi/\sqrt{3}) (\R0/ \RM)^{2}$. For AOT micelles on a 610 nm radius particle, this expression yields $\NP \approx 5 \times 10^{5}$, which is considerably larger than the particle charge
so the experimental system is clearly in the low-$\Theta$ limit. The discussion of section~\ref{sec-low-coverage} reveals that, for low-$\Theta$, the surface potential is controlled by the charging coefficient $\alpha$ (eq.~\ref{eq alpha}). In the limit when $\alpha \gtrsim 10^{2}$ the potential should approaches the saturated limit $\Ps$ and the charging mechanism becomes particularly simple. To identify if this is the case here, we estimate $\alpha$. Under the conditions of the electrokinetic experiments ($\vol > 10^{-4}$, $\volc = 3 \times 10^{-5}$) the number of micelles per particle is $> 5 \times 10^{8}$. For AOT, the equilibrium constant $\KN$ for adsorption of negative micelles is of the order of $\exp (- \b \GM )$ and $\KP \sim \exp (- \b \GM +2 \PH)$, from eq.~\ref{eq phi plateau}, which implies that $\alpha \sim 10^{7}$. Consequently we expect to be in the saturated potential limit where $\PH = \Ps$. This is consistent with the experimental observations -- namely that: (a) the measured potential $\PH$ does not depend on the concentration of reverse micelles (fig.~\ref{fig_zeta_aotzr}); (b) the equality of the charge and radius polydispersities seen in fig.~\ref{fig_poly_aotzr}; and (c) that the colloid charge is proportional to $\R0$  -- trends predicted by the charge regulation model introduced in sec.~\ref{sec-low-coverage}. Interpreting the measured surface potentials using Eq.~\ref{eq phi plateau} gives values for the free energy difference $\b(\GN - \GP)$ for the adsorption of positive and negatively-charged micelles of $-5.4\; \pm$ 0.1 for AOT, $+ 6.4\; \pm$ 0.4 for Zr(Oct)$_{2}$ and 0.3 $\pm$ 0.3 for the PHSA-PMMA copolymer. The zero value, within error, for PHSA-PMMA is very reasonable since the micelles are chemically identical to the colloid stabilizing layer and so we would expect no surface absorption. The preferential adsorption on PMMA of negative micelles, in the case of AOT, and positive micelles, in the Zr(Oct)$_{2}$ system, probably reflects the different hydrophobicities of the anionic AOT and cationic Zr(Oct)$_{2}$ surfactants.

\section{Conclusions} \label{sec-conclude}

Micrometer-sized  colloids may be charged in non-polar solvents, despite their ultra-low dielectric constant, by either the adsorption of ionic species or the dissociation of surface groups. The particle charge in a model system of sterically-stabilized polymer colloids has been determined using single particle optical microelectrophoresis (SPOM). A key advantage of SPOM is that it yields accurate measurements of the extremely small mobilities of isolated \textit{individual} colloidal particles typical of a non-polar solvent such as dodecane. This technique yields information on the
distribution of particle mobilities rather than simply recording the
average mobility, as provided by most conventional electrokinetic
techniques.  The effect of particle size, the nature and concentration of the reverse micelles on the magnitude and sign of the colloid charge distribution
has been investigated.  We find surprisingly simple charging characteristics, in stark contrast with the rather complex behavior often reported for charging in non-polar media (see for instance \cite{4437}). In an extensive study, using three different reverse micellar systems and five differently-sized particles, we find that the sign of the particle charge is determined by the chemical nature of the reverse micellar system. Interestingly, however the magnitude of the charge was found to be unaffected by the concentration of micelles and to scale linearly with the colloid radius. These generic features suggest a common mechanism of charging operates in micelle containing dispersions.

To interpret our data, we suggest colloids charge by the competitive adsorption of oppositely-charged reverse micelles. Within this model, the net charge $\Z0$ is determined by the \textit{difference} in the number of positive and negative micelles absorbed onto the particle surface. The composition of the surface layer depends non-linearly on the number of micelles in solution because oppositely-charged micelles are preferentially attracted to the charged colloid. We analyze a simple equilibrium model of this competitive adsorption process and show that, with increasing micelle concentration, the net charge rapidly saturates to a value $\Z0 = \Zs$. The saturated charge $\Zs$ scales linearly with the ratio of the particle radius to the Bjerrum length, $\Zs = \PH^{*} \R0 / \lb$, at screening lengths characteristic of non-polar suspensions. The coefficient in this expression, the saturated surface potential $\PH^{*}$, is the difference in the free energies of adsorption for negative and positive micelles in units of $\kb T$, $\PH^{*} =   \b( \GN - \GP) / 2$.  Analysis of our experimental data using this model gives free energy differences of order 5--6 $\kb T$, which look reasonable. Finally we note that that the model provides a coherent framework to understand and manipulate the charging of colloids in apolar solvents which will be highly beneficial for the future design of novel materials.

\section*{Acknowledgments}

It is a pleasure to thank W. Frith, and S. Clarke for helpful discussions. We thank S. Man Lam for help with the conductivity measurements. T. Narayanan, A. Mousaiid, and E. Di Cola are thanked for their help with the SAXS measurements. This research was sponsored by Unilever PLC and the UK Engineering and Physical Sciences Research Council.

\newpage


\begin{thebibliography}{10}

\bibitem{4951}
van~der Minne,~J.; Hermanie,~P. \emph{J. Coll. Sci.} \textbf{1951}, \emph{7},
  600--615.

\bibitem{3762}
Chen,~Y.; Au,~J.; Kazlas,~P.; Ritenour,~A.; Gates,~H.; McCreary,~M.
  \emph{Nature} \textbf{2003}, \emph{423}, 136--136.

\bibitem{3772}
Jo,~G.~R.; Hoshino,~K.; Kitamura,~T. \emph{Chem. Mater.} \textbf{2002},
  \emph{14}, 664--669.

\bibitem{3444}
Comiskey,~B.; Albert,~J.~D.; Yoshizawa,~H.; Jacobson,~J. \emph{Nature}
  \textbf{1998}, \emph{394}, 253--255.

\bibitem{5052}
Hao,~T. \emph{Adv. Mater.} \textbf{2001}, \emph{13}, 1847--1857.

\bibitem{4576}
Jones,~S.~A.; Martin,~G.~P.; Brown,~M.~B. \emph{J. Pharm. Sci.} \textbf{2006},
  \emph{95}, 1060--1074.

\bibitem{Morrison-868}
Morrison,~I.~D. \emph{Coll. Surf.} \textbf{1993}, \emph{71}, 1--37.

\bibitem{3480}
Croucher,~M.~D.; Lok,~K.~P.; Wong,~R.~W.; Drappel,~S.; Duff,~J.~M.;
  Pundsack,~A.~L.; Hair,~M.~L. \emph{J. Appl. Polym. Sci.} \textbf{1985},
  \emph{30}, 593 -- 607.

\bibitem{4939}
Pearlstine,~K.; Page,~L.; Elsayed,~L. \emph{J. Imag. Sci.} \textbf{1991},
  \emph{35}, 55--58.

\bibitem{4436}
Van Der~Hoeven,~P.~C.; Lyklema,~J. \emph{Adv. Colloid Interface Sci.}
  \textbf{1992}, \emph{42}, 205--277.

\bibitem{4949}
Leon,~O.; Rogel,~E.; Torres,~G.; Lucas,~A. \emph{Petroleum Sci. Technol.}
  \textbf{2000}, \emph{18}, 913--927.

\bibitem{3775}
Ryoo,~W.; Dickson,~J.~L.; Dhanuka,~V.~V.; Webber,~S.~E.; Bonnecaze,~R.~T.;
  Johnston,~K.~P. \emph{Langmuir} \textbf{2005}, \emph{21}, 5914--5923.

\bibitem{5051}
Touchard,~G. \emph{J. Electrostatics} \textbf{2001}, \emph{51-52}, 440--447.

\bibitem{4948}
Tolpekin,~V.~A.; van~den Ende,~D.; Duits,~M. H.~G.; Mellema,~J. \emph{Langmuir}
  \textbf{2004}, \emph{20}, 8460--8467.

\bibitem{3738}
Leunissen,~M.~E.; Christova,~C.~G.; Hynninen,~A.-P.; Royall,~C.~P.;
  Campbell,~A.~I.; Imhof,~A.; Dijkstra,~M.; Roij,~R.~v.; Blaaderen,~A.~v.
  \emph{Nature} \textbf{2005}, \emph{437}, 235--239.

\bibitem{3958}
Bartlett,~P.; Campbell,~A.~I. \emph{Phys. Rev. Lett.} \textbf{2005}, \emph{95},
  128302.

\bibitem{4981}
Parsegian,~A. \emph{Nature} \textbf{1969}, \emph{221}, 844--846.

\bibitem{4953}
Lyklema,~J.
\newblock \emph{Fundamentals of Interface and Colloid Science. Volume II:
  Solid-Liquid Interfaces};
\newblock Academic Press: London, 1995.

\bibitem{4954}
Parfitt,~G.~D.; Peacock,~J. Stability of Colloidal Dispersions in Nonaqueous
  Media. In \emph{Surface and Colloid Science: Volume 10.}; Matijevic,~E., Ed.;
\newblock Surface and Colloid Science;
\newblock Plenum Press: New York, 1978;
\newblock pp 163--226.

\bibitem{4955}
Kitahara,~A. Nonaqueous systems. In \emph{Electrical Phenomena at Interfaces:
  Fundamentals, Measurements and Applications}; Ohshima,~H., Furusawa,~K.,
  Eds., 2nd ed.;
\newblock Marcel Dekker: New York, 1998;
\newblock Chapter~4.

\bibitem{3771}
Hsu,~M.~F.; Dufresne,~E.~R.; Weitz,~D.~A. \emph{Langmuir} \textbf{2005},
  \emph{21}, 4881--4887.

\bibitem{3305}
Briscoe,~W.~H.; Horn,~R.~G. \emph{Langmuir} \textbf{2002}, \emph{18},
  3945--3956.

\bibitem{3768}
McNamee,~C.~E.; Tsujii,~Y.; Matsumoto,~M. \emph{Langmuir} \textbf{2004},
  \emph{20}, 1791--1798.

\bibitem{4437}
Kitahara,~A.; Amano,~M.; Kawasaki,~S.; Kon-no,~K. \emph{Colloid Polym. Sci.}
  \textbf{1977}, \emph{255}, 1118--1121.

\bibitem{4908}
Smith,~P.~G.; Patel,~M.~N.; Kim,~J.; Milner,~T.~E.; Johnston,~K.~P. \emph{J.
  Phys. Chem. C} \textbf{2007}, \emph{111}, 840--848.

\bibitem{3528}
Kitahara,~A.; Satoh,~T.; Kawasaki,~S.; Kon-No,~K. \emph{J. Coll. Interf. Sci.}
  \textbf{1982}, \emph{86}, 105--110.

\bibitem{3770}
Keir,~R.~I.; Suparno,; Thomas,~J.~C. \emph{Langmuir} \textbf{2002}, \emph{18},
  1463--1465.

\bibitem{4697}
Roberts,~G.~S.; Wood,~T.~A.; Frith,~W.~J.; Bartlett,~P. \emph{J. Chem. Phys.}
  \textbf{2007}, \emph{126}, 194503.

\bibitem{Antl-63}
Antl,~L.; Goodwin,~J.~W.; Hill,~R.~D.; Ottewill,~R.~H.; Owens,~S.~M.;
  Papworth,~S.; Waters,~J.~A. \emph{Colloids Surf.} \textbf{1986}, \emph{17},
  67--78.

\bibitem{304}
Nave,~S.; Eastoe,~J.; Penfold,~J. \emph{Langmuir} \textbf{2000}, \emph{16},
  8733--8740.

\bibitem{3449}
Kotlarchyk,~M.; Huang,~J.~S.; Chen,~S.~H. \emph{J. Phys. Chem.} \textbf{1985},
  \emph{89}, 4382 -- 4386.

\bibitem{3304}
Keir,~R.~I.; Watson,~J.~N. \emph{Langmuir} \textbf{2000}, \emph{16}, 7182.

\bibitem{295}
Papworth,~S. Ph.D.\ thesis, University of Bristol, 1993.

\bibitem{Bergenholtz-866}
Bergenholtz,~J.; Romagnoli,~A.~A.; Wagner,~N.~J. \emph{Langmuir} \textbf{1995},
  \emph{11}, 1559--1570.

\bibitem{3112}
O'Brien,~R.~W.; White,~L.~R. \emph{J. Chem. Soc., Faraday Trans. II}
  \textbf{1978}, \emph{74}, 1607.

\bibitem{3312}
Eicke,~H.~F.; Borkovec,~M.; Das-Gupta,~B. \emph{J. Phys. Chem.} \textbf{1989},
  \emph{93}, 314 -- 317.

\bibitem{4427}
Hall,~D.~G. \emph{J. Phys. Chem.} \textbf{1990}, \emph{94}, 429 -- 430.

\bibitem{4943}
Kallay,~N.; Chittofrati,~A. \emph{J. Phys. Chem.} \textbf{1990}, \emph{94},
  4755--4756.

\bibitem{4942}
Kallay,~N.; Tomic,~M.; Chittofrati,~A. \emph{Colloid Polym. Sci.}
  \textbf{1992}, \emph{270}, 194--196.

\bibitem{3451}
Bordi,~F.; Cametti,~C. \emph{Colloid Polym. Sci.} \textbf{1998}, \emph{276},
  1044--1049.

\bibitem{4433}
Hansen,~S. \emph{J. Chem. Phys.} \textbf{2004}, \emph{121}, 9111--9115.

\bibitem{4432}
Tirado,~M.~M.; Torre,~J.~G. \emph{J. Chem. Phys.} \textbf{1979}, \emph{71},
  2581--2587.

\bibitem{3212}
Thomas,~J.~C.; Crosby,~B.~J.; Keir,~R.~I.; Hanton,~K.~L. \emph{Langmuir}
  \textbf{2002}, \emph{18}, 4243--4247.

\bibitem{4230}
Garbow,~N.; Eyers,~M.; Palberg,~T.; Okubo,~T. \emph{J. Phys.: Condens. Matter}
  \textbf{2004}, \emph{16}, 3835--3842.

\bibitem{4438}
Strubbe,~F.; Beunis,~F.; Neyts,~K. \emph{J. Coll. Interf. Sci.} \textbf{2006},
  \emph{301}, 302--309.

\bibitem{3218}
Aubouy,~M.; Trizac,~E.; Bocquet,~L. \emph{J. Phys. A: Math. Gen.}
  \textbf{2003}, \emph{36}, 5835--5840.

\bibitem{3206}
Alexander,~S.; Chaikin,~P.~M.; Grant,~P.; Morales,~G.~J.; Pincus,~P. \emph{J.
  Chem. Phys.} \textbf{1984}, \emph{80}, 5776--5781.

\bibitem{5028}
Manne,~S.; Gaub,~H.~E. \emph{Science} \textbf{1995}, \emph{270}, 1480--1482.

\bibitem{4934}
Krishnakumar,~S.; Somasundaran,~P. \emph{Coll. Surf.} \textbf{1996},
  \emph{117}, 227--233.

\bibitem{4896}
Chan,~D.; Perram,~J.~W.; White,~L.~R.; Healy,~T.~W. \emph{J. Chem. Soc.,
  Faraday Trans. 1,} \textbf{1975}, \emph{71}, 1046 -- 1057.

\bibitem{4881}
Strubbe,~F.; Beunis,~F.; Marescaux,~M.; Neyts,~K. \emph{Phys. Rev. E}
  \textbf{2007}, \emph{75}, 031405.

\bibitem{4987}
Lemaire,~B.; Bothorel,~P.; Roux,~D. \emph{J. Phys. Chem.} \textbf{1983},
  \emph{87}, 1023--1028.

\bibitem{4966}
Pincus,~P.~A.; Safran,~S.~A. \emph{J. Chem. Phys.} \textbf{1987}, \emph{86},
  1644--1645.

\bibitem{4986}
Cassin,~G.; Badiali,~J.~P.; Pileni,~M.~P. \emph{J. Phys. Chem.} \textbf{1995},
  \emph{99}, 12941--12946.

\end{thebibliography}
\end{document}